# Controllable coherent perfect absorption in a composite film

**Shourya Dutta-Gupta,**[1] **Olivier J. F. Martin,**[1,4] **S. Dutta Gupta,**[2,5] **and G. S. Agarwal**[3,6]

[1]*Nanophotonics and Metrology Laboratory,
Swiss Federal Institute of Technology (EPFL), Lausanne, CH-1015, Switzerland*
[2]*School of Physics,
University of Hyderabad, Hyderabad, 500019, India*
[3]*Department of Physics,
Oklahoma State University Stillwater, Oklahoma, 74078, USA*

[4]*olmatrin@epfl.ch*

[5]*sdghyderabad@gmail.com*

[6]*girish.agarwal@okstate.edu*

**Abstract:** We exploit the versatility provided by metal–dielectric composites to demonstrate controllable coherent perfect absorption (CPA) in a slab of heterogeneous medium. The slab is illuminated by coherent light from both sides, at the same angle of incidence and the conditions required for CPA are investigated as a function of the different geometrical parameters. The simultaneous realization of CPA at two distinct frequencies is also shown. Finally, our calculations clearly elucidate the role of absorption as a necessary prerequisite for CPA.

© 2011 Optical Society of America

## 1. Introduction

There has been recently a great deal of interest in critical coupling (CC) in micro and nano structures [1, 2, 3, 4, 5]. Critical coupling of incident electromagnetic radiation to a given micro– nano–structure refers to the case where the entire incident energy is absorbed in the structure, leading to null scattering (see for example Refs. [3, 4]). Initial research on critical coupling involved the coupled resonator optical waveguides (CROWs) proposed by Yariv [6]. A realization of such system was the coupled fiber–microsphere or the fiber–disc system [1]. The purpose here was to slow down light in order to stop and store it eventually. Later studies were directed to planar structures with a very thin layer of absorbing material on a distributed Bragg reflector (DBR) substrate [3, 4, 5]. A spacer layer between the absorbing layer and the DBR controlled the amplitude and phase of the waves reflected from the different interfaces. For light at normal incidence from the top (on absorbing layer side), the DBR ensured null transmission for waves with frequency in the rejection band of the structure. The spacer layer thickness was controlled such that all the reflected waves coming from different interfaces interfered destructively in the medium of incidence, leading to null reflection. Thus, the structure neither transmitted nor reflected, implying the critical coupling of the incident light to the structure. The entire incident light was "perfectly" absorbed by the only few–nanometer thick lossy layer. Physically this amounts to having a purely imaginary component of the Poynting vector normal to the surface.

The use of composites [7] and metamaterials [8, 9, 10] as the absorbing layer in CC structures opens new possibilities for tailoring their optical response [4, 5, 11, 12]. Composites, in particular metal–dielectric nano–composites, have found interesting applications in optics, especially since the eighties, when researchers appreciated that the effective medium properties of the composite can exceed those of its individual constituents [13, 14, 15]. The resulting enhancement in the local fields and local density of states can lead to interesting linear, nonlinear and cavity QED applications [15, 16, 17, 18]. The possibilities offered by such composites are even more striking for metal inclusions in a dielectric host, since localized plasmon resonances can be excited in the metallic nano–inclusions. Critical coupling at two distinct frequencies for both normal and oblique incidences was reported using such nano–composites [4, 5]. The splitting of the normal mode of the total absorption dip using intracavity resonant atoms for a cavity with metamaterial mirrors was also shown [18].

Usually a single coherent source illumination from one side is used in critical coupling. A generalization to two incident coherent beams from opposite sides leads to the so–called coherent perfect absorption (CPA), with the added flexibility that the two beams can be controlled separately [19, 20, 21, 22]. Due to complete destructive interference at both sides there can be perfect absorption of the incident beams, which is sometimes referred to as time reversed lasing (near threshold) [19, 20]. Recently, there have been several reports on CPA and related phenomena in different systems [21, 22]. Let us emphasize that the absence of scattering both in forward and backward directions originates from the same physical phenomenon for CC and CPA: destructive interference of two beams. The difference between CC and CPA resides in the implementation detail: two incident beams are used in CPA, while only one is used in CC where the second beam is generated by the structure itself.

In this paper we show that the concept of CPA can be extended to versatile geometries using a heterogeneous metal–dielectric composite layer as absorber. This approach provides great flexibility with respect to the material sample, illumination configuration, and frequency range where CPA can be realized. In a heterogeneous composite medium it is possible to tune the localized plasmon resonance and the resulting absorption and dispersion, by simply varying the inclusions volume fraction. Here we present a detailed study of CPA in a slab of composite medium with illumination by coherent waves at identical angle of incidence from both sides. Identical angle of incidence ensures the interference of the transmitted light, say, incident from the left of the medium, with the reflected one incident from right. Perfect destructive interference corresponds to the case when the amplitude of both waves are the same and their phase difference is $\pi$. If both incident waves are identical, this leads to perfect cancellation on both sides and CPA. We demonstrate various realizations and explore the possibility of having CPA at two distinct frequencies for the same structure. We also present results which are indicative of a critical minimum of absorption below which CPA is not possible. This clearly demonstrates the necessity of absorption for CPA.

## 2. Formulation of the problem

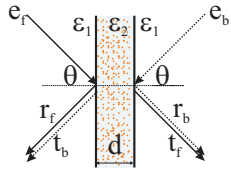

Fig. 1. Schematics of the CPA setup.

Consider the geometry shown in Fig. 1, where a metal–dielectric composite layer of thickness $d$ is excited by two coherent monochromatic waves with unit amplitude from both sides. All the media in Fig. 1 are assumed nonmagnetic and the angle of incidence $\theta$ is the same for both waves. For future reference we label the wave propagating forward and incident from left (right) and resulting reflected and transmitted waves with subscript $f$ ($b$). The structure is symmetric, since the medium of incidence and emergence are the same. Symmetry ensures that the total scattered amplitudes in the medium of incidence and emergence are the same since the reflected and transmitted amplitudes individually are the same ($r_f = r_b, t_f = t_b$). The nature of scattering in both directions is governed by the interference between, say, $r_f$ and $t_b$. It will be destructive and leading to CPA if the magnitudes of these waves are the same with a phase difference of $\pi$, i.e., $|r_f| = |t_b|$ and $|\Delta\phi| = |\phi_r - \phi_t| = \pi$, where $\phi_r$ and $\phi_t$ refer to the phases of $r_f$ and $t_b$, respectively. We have dropped the subscript in the phases and retained them in the amplitudes to stress that both waves are present in the same medium to interfere. Again the inherent symmetry implies that $r_f + t_b = r_b + t_f$, leading to the same scattered amplitudes on both sides. Thus destructive cancellation in the incidence medium would imply the same in the emergence medium, leading to CPA.

The complex reflection and transmission amplitudes for any given polarization for the structure shown in Fig. 1 can be easily calculated using the standard characteristic matrix approach [23]. Linearity of the system can be exploited to calculate the contributions for left and right incidences separately and superpose them to obtain the scattered amplitudes on both the sides. As previously mentioned, localized plasmon resonances in the composite play a key role the scattering in the system. The optical properties of this metal–dielectric medium are obtained from the Bruggeman formula, where both constituents in the two–component medium are treated on the same footing [10]. The permittivity of the composite is thus given by

$$\varepsilon_2 = \frac{1}{4}\left\{(3f_m - 1)\varepsilon_m + (3f_d - 1)\varepsilon_d \pm \sqrt{[(3f_m - 1)\varepsilon_m + (3f_d - 1)\varepsilon_d]^2 + 8\varepsilon_m\varepsilon_d}\right\}, \quad (1)$$

where $f_m$ and $\varepsilon_m$ ($f_d$ and $\varepsilon_d$) are the volume fraction and the permittivity of the metal (dielectric), respectively. Since the composite has two components, we have $f_d = 1 - f_m$. The square root is taken such that the imaginary part of the permittivity is positive to ensure causality [24].

The flexibility offered by the structure in Fig. 1 in the context of CPA can easily be assessed, even before any calculations. There are now several parameters controlling the nature of the scattered light, namely, the width $d$, the angle of incidence $\theta$; both controlling the optical path and hence the single–pass or roundtrip phase and attenuation for a given frequency. Most importantly, the volume fraction of both constituents of the metal–dielectric composite provide a very efficient control of its absorption and dispersion. In the next section we demonstrate this freedom of choice of optical parameters for a gold–silica composite. We consider very low volume fraction for the metal, leading to a localized plasmon resonance around $\lambda = 530$ nm and producing CPA dips near $\lambda = 560$ nm. Figure 2 shows the dependence of the real and imaginary parts of the permittivity for a gold–silica composite as a functions of the wavelength $\lambda$ for two different values of the metal volume fraction: $f_m = 0.002$ and $f_m = 0.08$. The dramatic changes of the absorption and localized plasmon resonance wavelength are clearly visible. An increased value of $f_m$, say $f_m = 0.08$, red–shifts the localized plasmon modes to about $\lambda = 630$ nm, bringing the CPA effect within the reach of laser diodes in the $\lambda = 750$–$790$ nm spectral range.

## 3. Results and Discussion

All the calculations were performed for a gold–silica composite layer illuminated by a transverse electric (TE) polarized plane waves incident at an angle $\theta = 45°$. The case for TM–polarization is also straightforward, yielding analogous results but with the CPA condition being satisfied for different parameters (not shown here). The dielectric function of gold $\varepsilon_m$ was obtained by interpolating the experimental data by Johnson and Christy [25]. Other parameters were taken as follows: $\varepsilon_1 = 1.0$, $\varepsilon_d = 2.25$. In order study the influence of the geometrical and material parameters, we have computed the modulus of the reflected and transmitted amplitudes, $|r_f|$, $|t_b|$, their phase difference $\Delta\phi$ and the intensity of the scattered light $|r_f + t_b|^2$ as functions of $d$, $f_m$ or $\lambda$.

As previously mentioned, the CPA results from an extremely delicate balance where the interfering waves have the same modulus and differ by a phase of $\pi$. This is investigated in

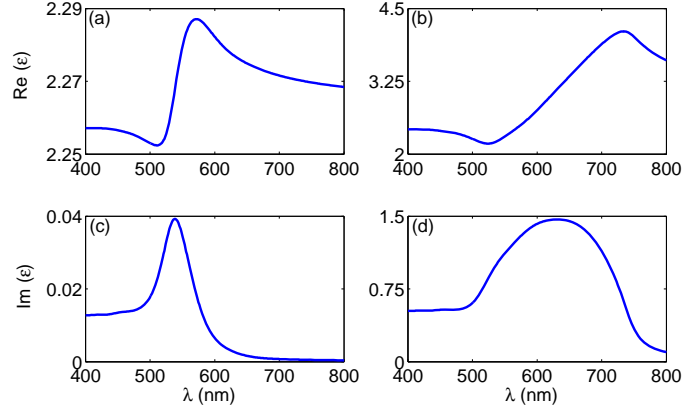

Fig. 2. (a), (b) Real and (c), (d) imaginary parts of the dielectric constant for a gold–silica composite at two different volume fractions: (a), (c) $f_m = 0.002$ and (b), (d) $f_m = 0.08$. Here $\varepsilon_d = 2.25$ and $\varepsilon_m$ is taken from the work of Johnson and Christy [25].

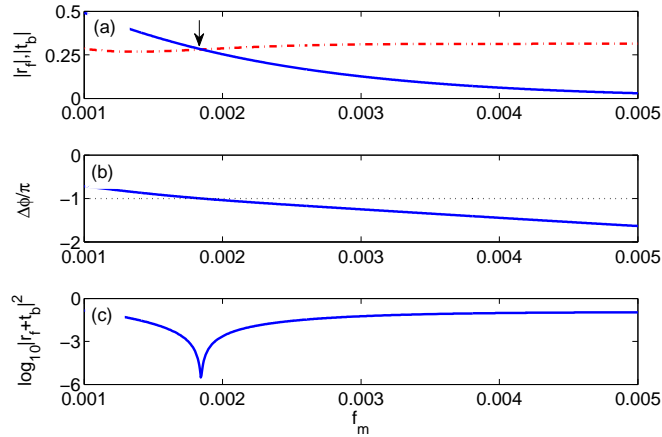

Fig. 3. (a) Absolute values of reflected (dashed line) and transmitted (solid line) amplitudes $|r_f|$ and $|t_b|$ (arrow shows the point where conditions for CPA are satisfied), (b) phase difference $\Delta\phi$ between the forward reflected and backward transmitted plane waves and (c) $log_{10}|r_f + t_b|^2$, as a function of $f_m$ for $d=10\,\mu$m and $\lambda = 556$nm.

Fig. 3, where the amplitude of the reflected and transmitted waves are shown as a function of $f_m$ for $\lambda = 556$nm and $d = 10\,\mu$m. A logarithmic scale is used in Fig. 3(c) to highlight the depth of the dip where CPA occurs for $f_m = 0.001844$, where the required conditions are fulfilled, i.e., $|r_f| = |t_b|$ and $|\Delta\phi| = \pi$.

The dependence of $|r_f|$, $|t_b|$, $|\Delta\phi|$ and $log_{10}|r_f + t_b|^2$ as functions of $\lambda$ for a composite layer with $f_m = 0.001844$ and $d = 10\,\mu$m is shown in Fig. 4. These data confirm the necessary conditions for CPA, i.e., destructive interference between $r_f$ and $t_b$. They also indicate new possibilities, such as having CPA at two distinct frequencies for the same structure under illumination at the same angle of incidence. Note in Fig. 4(a) the multiple crossing of the $|r_f|$ and $|t_b|$ curves, implying amplitude matching at more than one wavelength. While the phase difference vanishes at one of these crossing (at $\lambda = 556$nm), it is slightly off for the other wavelengths. It is thus expected that in the parameter space a nearby point can be found, where the matching is complete atleast at two distinct frequencies. This is indeed possible and the results are shown in Fig. 5 for $f_m = 0.001618$ and $d = 9.909\,\mu\mathrm{m}$. It is not ruled out that a judicious choice of parameters, possibly with periodic variations of the refractive index, could lead to CPA at more than two frequencies.

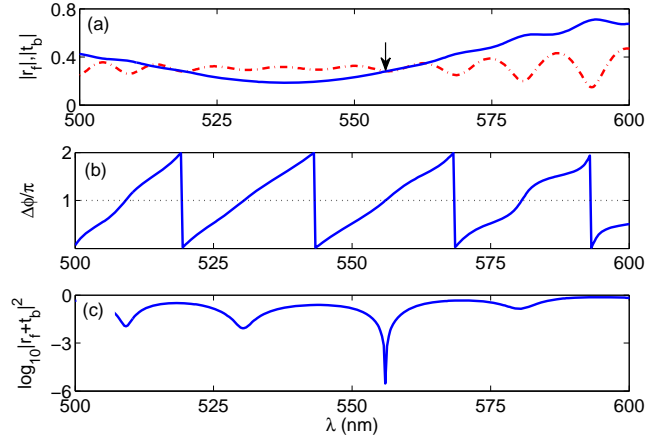

Fig. 4. Same as in Fig. 3 for $f_m = 0.001844$ and $d = 10\,\mu\mathrm{m}$.

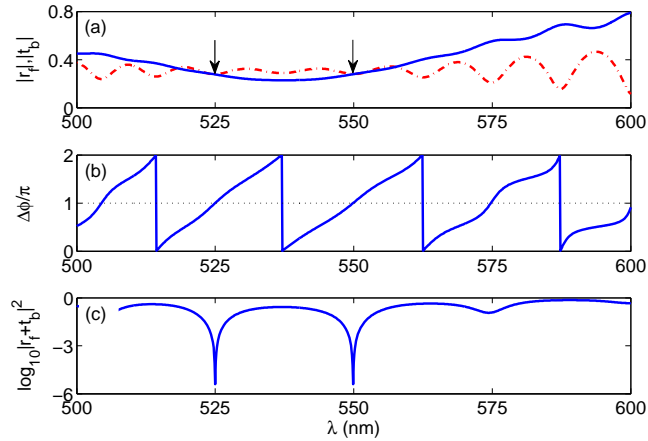

Fig. 5. Same as in Fig. 3 for $f_m = 0.001618$ and $d = 9.909\,\mu\mathrm{m}$.

In order to further explore the influence of the different geometrical parameters on the CPA condition, we study in Fig. 6 the scattered intensity $log_{10}|r_f + t_b|^2$ as a functions of two of the parameters (keeping the third one constant) in a three–parameter space (comprising $f_m$, $\lambda$ and $d$). The results for the log of the scattered intensity on one side are shown as color plots as a function of $f_m$ and $\lambda$ for $d = 10\,\mu\mathrm{m}$ Fig. 6(a), as a function of $d$ and $\lambda$ for $f_m = 0.0015$

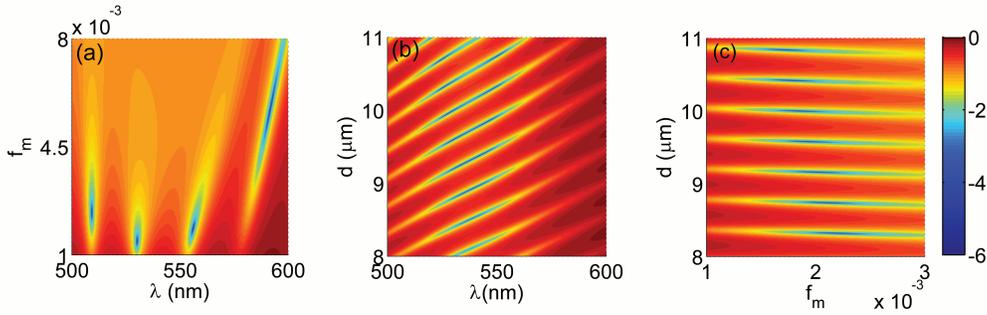

Fig. 6. Color map of $log_{10}|r_f+t_b|^2$ as functions of (a) $f_m$ and $\lambda$ for $d = 10\,\mu$m, (b) $d$ and $\lambda$ for $f_m = 0.0015$, and (c) $d$ and $f_m$ for $\lambda = 556$.

Fig. 6(b), and as a function of $d$ and $f_m$ for $\lambda = 556$ nm Fig. 6(c). The islands and stripes in this figure indicate that it is nontrivial to meet the CPA condition, which results from a very delicate balance ensuring destructive interference. As can be seen from Fig. 6(b) and (c), the stripes contain the islands and the CPA condition is met at a point near the middle of the island. The occurrence of several islands with sharp dips suggests that CPA is possible for several pairs of parameter values, when the third parameter is kept fixed. The dependence shown in Figs. 6(a) and (b) is also indicative of another important fact, namely, absorption is essential for CPA. Fig. 6(a) clearly points to a limiting volume fraction of metal, below which CPA is ruled out. Note that in our model the metallic inclusions are the only source of losses and hence existence of a minimum critical value of metal volume fraction implies a similar minimum absorption for the realization of CPA. Both Figs. 6(a) and (b) suggest that a possible adjustment of $d$ and $f_m$, respectively, can lead to two 'point dips' on the same horizontal line, thereby leading to CPA at two frequencies for the same structure. Finally, Fig.6(c) implies that an increase of $d$, resulting in an overall absorption increase, can lead to CPA at a lower volume fraction $f_m$.

## 4. Conclusion

We have studied CPA in a two–component metal–dielectric heterogeneous medium illuminated from both sides under plane wave oblique incidence. The composite medium permittivity was approximated by the effective medium theory of Bruggeman and it was shown that the flexibility of the dielectric response of such composite medium, which is due to the localized plasmon resonance and can be controlled by the metal volume fraction of metal, provides control on the CPA conditions and enable its realization over a large range of optical wavelengths. This flexibility was further exploited to obtain CPA a two different frequencies for the same structure. A detailed analysis of the CPA scattering dips as a function of the different geometrical parameters of the system was carried out. Our data clearly demonstrate that absorption is a necessary prerequisite for CPA. These results have far reaching implications for sensing and other optical devices.

## Acknowledgements

One of the authors (SDG) is thankful to the Department of Science and Technology, Government of India, for partially supporting this work. This work was supported by the Swiss National Science Foundation (Grant No. CR23I2_130164).